\documentclass[final]{cvpr}

\usepackage{times}
\usepackage{epsfig}
\usepackage{graphicx}
\usepackage{amsmath}
\usepackage{amssymb} 

\usepackage{multirow}
\usepackage{listings}

\usepackage[normalem]{ulem}
\useunder{\uline}{\ul}{}

\usepackage{arydshln}
\usepackage{paralist}

\usepackage{cuted}
\usepackage{float}
\usepackage{placeins}
\usepackage{algorithm}
\usepackage{algpseudocode}
\usepackage{varwidth}
\usepackage{microtype}

\usepackage[pagebackref=true,breaklinks=true,colorlinks,bookmarks=false]{hyperref}

\def\cvprPaperID{0004} 
\def\confYear{CVPR 2021}
\begin{document}

\title{End-to-end optimized image compression with competition of prior distributions}

\author{Benoit Brummer\\
intoPIX\\
Mont-Saint-Guibert, Belgium\\
{\tt\small b.brummer@intopix.com}
\and
Christophe De Vleeschouwer\\
Université catholique de Louvain\\
Louvain-la-Neuve, Belgium\\
{\tt\small christophe.devleeschouwer@uclouvain.be}
}

\maketitle

%
%
%

\begin{abstract}

Convolutional autoencoders are now at the forefront of image compression research. To improve their entropy coding, encoder output is typically analyzed with a second autoencoder to generate per-variable parametrized prior probability distributions. We instead propose a compression scheme that uses a single convolutional autoencoder and multiple learned prior distributions working as a competition of experts. Trained prior distributions are stored in a static table of cumulative distribution functions. During inference, this table is used by an entropy coder as a look-up-table to determine the best prior for each spatial location. Our method offers rate-distortion performance comparable to that obtained with a predicted parametrized prior with only a fraction of its entropy coding and decoding complexity.

\end{abstract}

\section{Introduction}
Image compression typically consists of a transformation step (including quantization) and an entropy coding step that attempts to capture the probability distribution of a transformed context to generate a smaller compressed bitstream. Entropy coding ranges in complexity from simple non-adaptive encoders \cite{jpeg,jpegxs-entropy} to complex arithmetic coders with adaptive context models \cite{cabac,jpegxl}. The entropy coding strategy has been revised to address the specificities of learned compression. More specifically, for recent works that make use of a convolutional autoencoder \cite{autoencoder} (AE) as the all-inclusive transformation and quantization step, 
the entropy coder relies on a cumulative probability model (CPM) trained alongside the AE \cite{balle2017}. This model estimates the cumulative distribution function (CDF) of each channel coming out of the AE and passes these learned CDFs to an entropy coder such as range encoding \cite{rangeencoding}.

Such a simple method outperforms traditional codecs like JPEG2000 but work is still needed to surpass complex codecs like BPG. Johannes Ballé et al. (2018) \cite{balle2018} proposed analyzing the output of the convolutional encoder with another AE to generate a floating-point scale parameter that differs for every variable that needs to be encoded by the entropy coder, thus for every location in every channel. This method has been widely used in subsequent works but introduces substantial complexity in the entropy coding step because a different CDF is needed to encode every variable in the latent representation of the image, whereas the single AE method by Ballé et al. (2017) \cite{balle2017} reused the same CDF table for every latent spatial location.

Our work uses the principle of competition of experts \cite{competitionOfExperts0,competitionOfExperts1} to get the best out of both worlds. 
Multiple prior distributions compete for the lowest bit cost on every spatial location in the quantized latent representation. During training, only the best prior distribution is updated in each spatial location, further improving the prior distributions specialization. CDF tables are fixed at the end of training. Hence, at testing, the CDF table resulting in the lowest bitcost is assigned to each spatial location of the latent representation. The rate-distortion (RD) performance obtained is comparable to that obtained with a parametrized distribution \cite{balle2018}, yet the entropy coding process is greatly simplified since it does not require a per-variable CDF and can build on look-up-tables (LUT) rather than the computation of analytical distributions.

\section{Background}

Entropy coders such as range encoding \cite{rangeencoding} require $\textit{\textbf{cdf}}$s where, for each variable to be encoded, the probability that a smaller or equal value appears is defined for every allowable value in the latent representation space. Johannes Ballé et al.'s seminal work (2017) \cite{balle2017} consists of an AE, computing a latent image representation consisting in $\text{C}_\text{L}$ channels of size $\text{H}_\text{L} \times \text{W}_\text{L}$, and a CPM, consisting of one CDF per latent output channel, which are trained conjointly. The latent representation coming out of the encoder is quantized then passed through the CPM. The CPM defines, in a parametrized and differentiable manner, a CDF per channel. At the end of training, the CPM is evaluated at every possible value\footnotemark[1] to generate the static CDF table. The CDF table is not differentiable, but going from a differentiable CPM to a static CDF table speeds up the encoding and decoding process. The CDF table is used to compress latent representations with an entropy coder, the approximate bit cost of a symbol is the binary logarithm of its probability.

Ballé et al. (2018) improved the RD efficiency by replacing the unique CDF table with a Gaussian distribution parametrized with a hyperprior (HP) sub-network \cite{balle2018}. The HP generates a scale parameter, and in turn a different CDF, for every variable to be encoded. Thus, complexity is added by exploiting the parametrized Gaussian prior during the entropy coding process, since a different CDF is required for each variable in the channel and spatial dimensions.  

 Minnen et al. proposed a scheme where one of multiple probability distributions is chosen to adapt the entropy model locally \cite{minnenmp}. However, these distributions are defined a posteriori, given the encoder trained with a global entropy model. Thus \cite{minnenmp} does not perform as well as the HP scheme \cite{balle2018} per \cite[Fig.~2a]{minnen}. In contrast, the present method jointly optimizes the local entropy models and the AE in an end-to-end fashion that results in greater performance. 
Minnen et al. \cite{minnen} later proposed to improve RD with the use of an autoregressive sequential context model. However, as highlighted in \cite{liujiaheng}, this is obtained at the cost of increased runtime by several orders of magnitude. 
Subsequent works have attempted to reduce complexity of the neural network architecture \cite{johnston} and to bridge the RD gap with Minnen's work \cite{liujiaheng}, but entropy coding complexity has remained largely unaddressed and has instead evolved towards increased complexity \cite{minnen,gmm,minnen2020} compared to \cite{balle2018}. The present work builds on Ballé et al. (2017) \cite{balle2017} and achieves the performance of Ballé et al. (2018) \cite{balle2018} without the complexity introduced by a per-variable parametrized probability distribution. We chose Ballé et al. (2017) as a baseline because it corresponds to the basic unit adopted as a common reference and starting point for most models proposed in the recent literature to improve compression quality \cite{balle2018,minnen,liujiaheng,minnen2020}. Due to its generic nature, our contribution remains relevant for the newer, often computationally more complex, incremental improvements on Ballé et al. (2017).


\section{Competition of prior distributions}

\begin{figure}
\centering
\includegraphics[width=\linewidth]{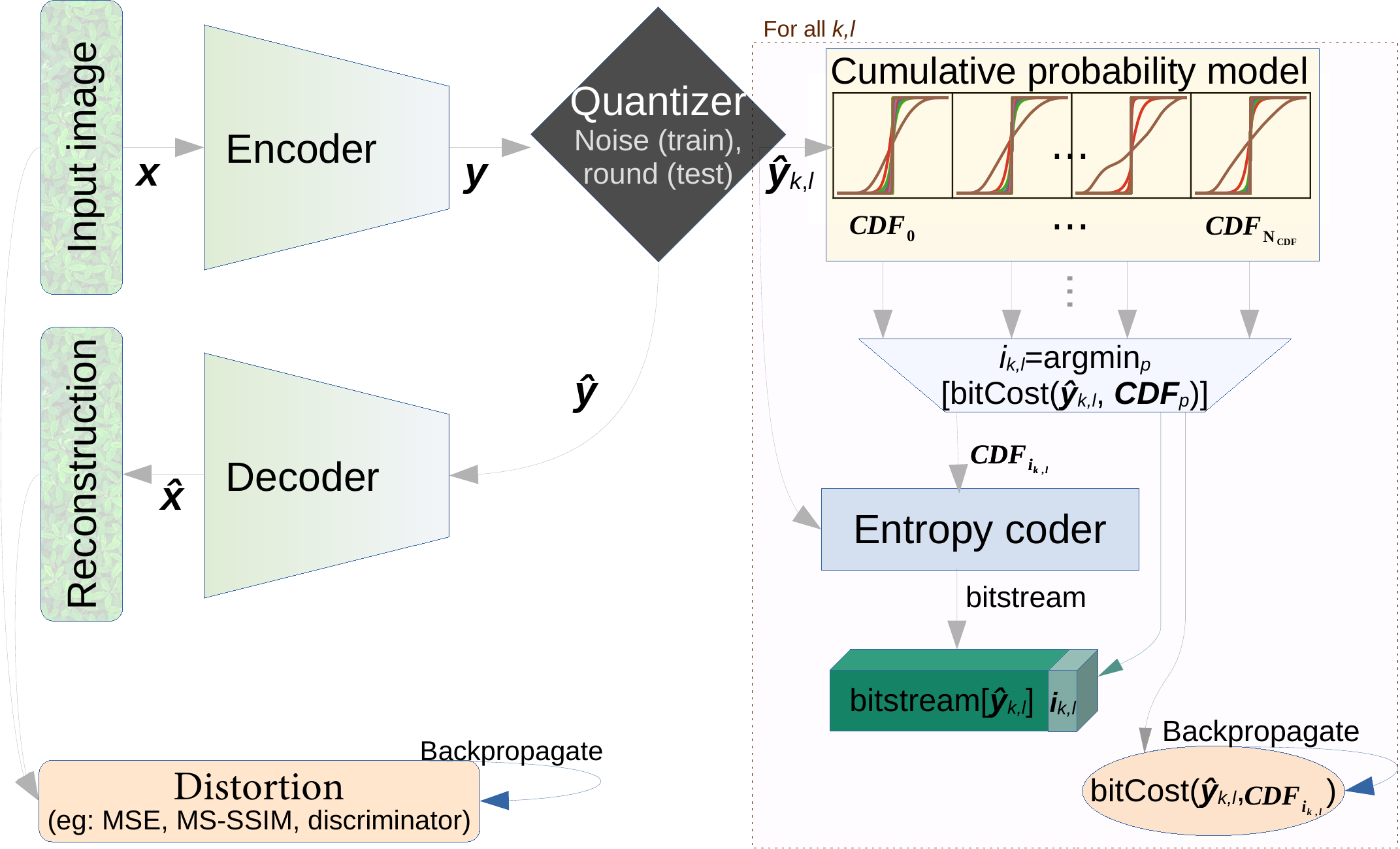}
\caption{AE compression scheme with competition of prior distributions. The AE architecture is detailed in \cite[Fig.~4]{balle2018}. The indices $\textit{\textbf{i}}$ denote the indices of CDF tables that minimizes the bitcount for each latent spatial dimension. Loss = Distortion + $\lambda$ bitCost.}
\label{fig:mparch}

\end{figure}

\begin{figure}
\centering
\includegraphics[width=\linewidth]{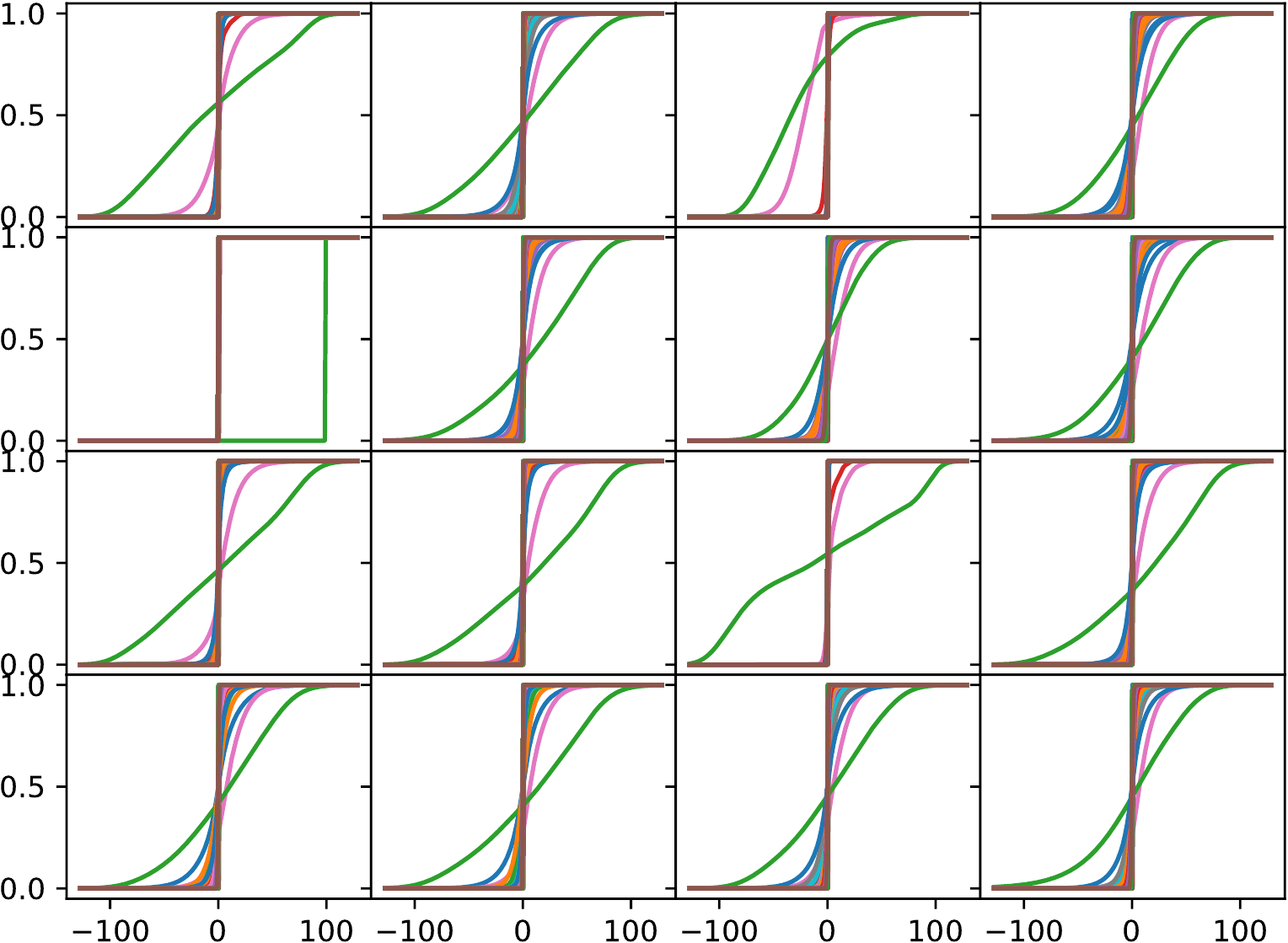}
\caption{We observe some diversity among the 16 cumulative distribution functions learned by a network trained with MSE loss and $\lambda =4096$. Each box presents a CDF table and each colored line corresponds to the $\textit{\textbf{cdf}}$ of one of 256 latent channels. The best fitting CDF table is selected for each latent spatial location.}
\label{fig:16priors}
\end{figure}

\begin{figure}
\centering
\includegraphics[width=\linewidth]{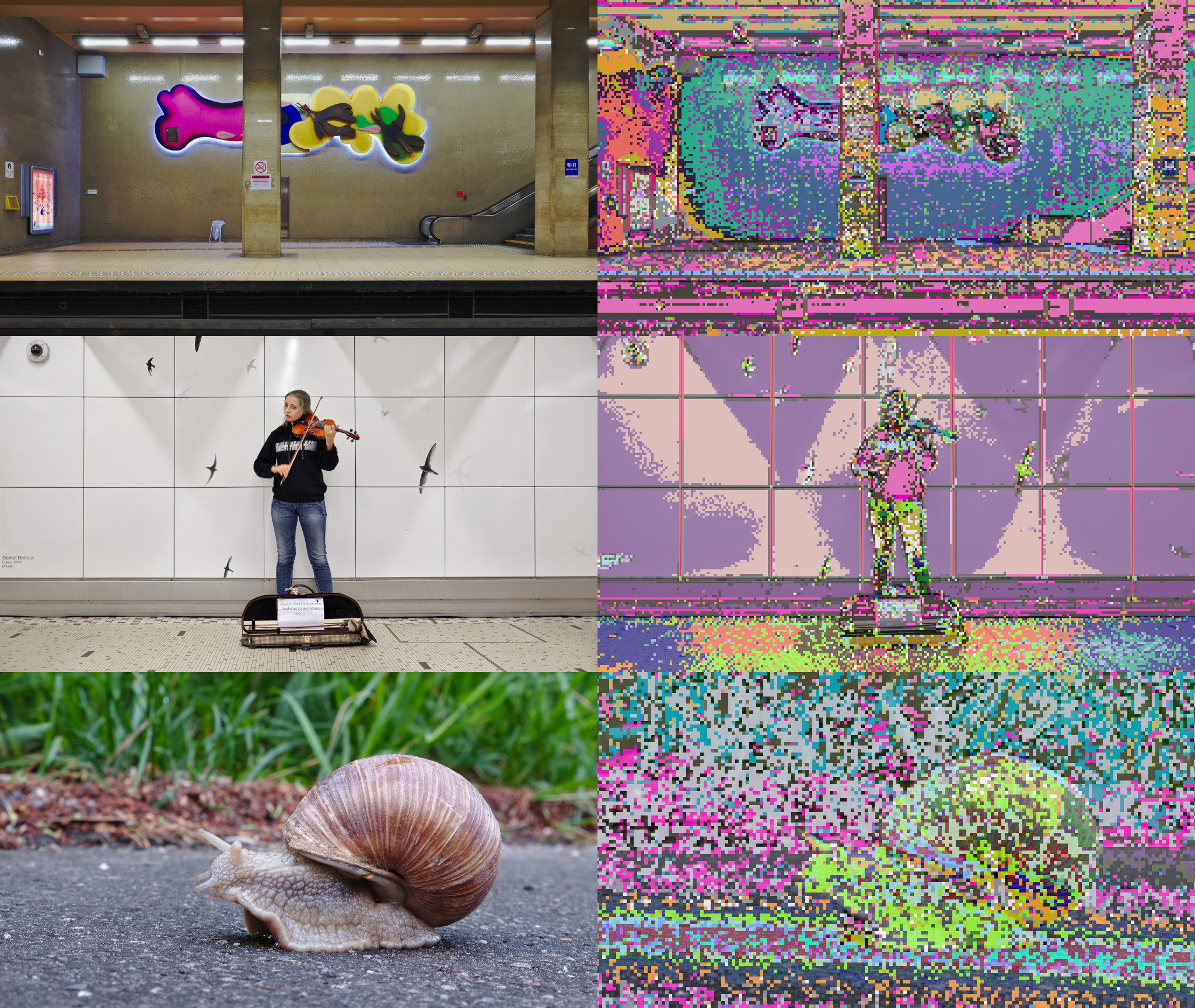}
\caption{Segmentation of three test images \cite{commonstestphotographs}: each distinct color represents one of 64 CDF tables used to encode a latent spatial location ($16\times 16$ pixels patch)}
\label{fig:segvis}
\end{figure}

Our proposed method introduces competitions of expert \cite{competitionOfExperts0,competitionOfExperts1} prior distributions: a single AE transforms the image and a set of prior distributions are trained to model the $\textit{\textbf{CDF}}$ of the latent representation in each spatial location. For each latent spatial dimension the CDF table which minimizes bit cost is selected; that prior is either further optimized on the features it won in the training mode, or its index is stored for decoding in the inference mode. This scheme is illustrated in \autoref{fig:mparch}, a set of 16 optimized CDF tables is shown in \autoref{fig:16priors}, and three sample images are segmented by ``winning'' CDF table in \autoref{fig:segvis}.

All prior distributions are estimated in parallel by considering $\text{N}_\text{CDF}$ CDF tables, and selecting, as a function of the encoded latent spatial location, the one that minimizes the entropy coder bitcount. The CDF table index is determined for each spatial location by evaluating each CDF table in inference. This can be done in a vectorized operation given sufficient memory. During training the CPM is evaluated instead of CDF tables such that the probabilities are up to date and the model is differentiable, and the bit cost is returned as it contributes to the loss function. 
%
%
The cost of CDF table indices has been shown to be neglectable due to the reasonably small number of priors, which in turns results from the fact that little gain in latent code entropy has been obtained by increasing the number of priors.

In all our experiments , the AE architecture follows the one in Ballé et al. (2018) \cite{balle2018}, without the HP, since we found that the AE from \cite{balle2018} offers better RD than the one described in Ballé et al. (2017) \cite{balle2017}, even with a single CDF table. A functional training loop is described in \autoref{eq:trainloop}. 

\algrenewcommand\algorithmicindent{0.5em}%
\begin{algorithm}
\small
\caption{Training loop}
\label{eq:trainloop}
\begin{algorithmic}
\State $\textit{\textbf{y}}  \gets $ model.Encoder($\textit{\textbf{x}}$)
\State $\textit{\textbf{\^{y}}} \gets$ quantize($\textit{\textbf{y}}$)
\State $\textit{\textbf{\^{x}}} \gets $ clip(model.Decoder($\textit{\textbf{\^{y}}}$), 0, 1)
\State \textit{distortion} $\gets$ visualLossFunction($\textit{\textbf{\^{x}}}$, $\textit{\textbf{x}}$)
\For{$0 \leq k<\text{H}_\text{L} \text{ and } 0 \leq l<\text{W}_\text{L}$}
\State $\textit{\textbf{bitCost}}\left[k,l\right] \gets  \min_{\textit{i}<\text{N}_\text{CDF}}$\par\hskip\algorithmicindent$\big|-\log_2\big(\textit{\textbf{CPM}}_\textit{i}(\textit{\textbf{\^{y}}}[k,l]+0.5) - \textit{\textbf{CPM}}_\textit{i}(\textit{\textbf{\^{y}}}[k,l]-0.5)\big)\big|$
\EndFor\Comment{\footnotesize\textit{\textbf{CPM}} is the differentiable version of \textbf{\textit{CDF}}}\normalsize
\State $\textit{Loss} \gets \textit{distortion} \times \lambda + \left| \textit{\textbf{bitCost}} \right|$
\State $\textit{Loss}$.backward()\\
\end{algorithmic}

\end{algorithm}


\section{Experiments}

\subsection{Method}\label{sec:methods}

These experiments are based on the PyTorch implementation of Ballé et al. (2018) \cite{balle2018} published by Liu Jiaheng \cite{ptcompression,liujiaheng}. To implement our proposed method, the HP is omitted in favor of competition of expert prior distributions. The CPM is that defined in \cite{ptcompression} with an additional $\text{N}_\text{CDF}$ dimension to compute all CDF tables in parallel. Theoretical results are verified using the $\text{torchac}$ range coder \cite{torchac,torchac-code,rangeencoding}. A functional training loop is described in \autoref{eq:trainloop}, and source code is provided on \url{https://github.com/trougnouf/Manypriors}. To ensure that all priors get an opportunity to train, the prior distributions that have not been used for at least fifty steps are randomly assigned to spatial locations with largest bitcounts, to be forced to train.
The Adam optimizer \cite{adam} is used with a starting learning rate (LR) of 0.0001 for the AE and 0.001 for the CPM. Performance is tested every 2500 steps in inference mode on the validation set, and the LR 
is decayed by a factor of 0.99 if 
the performance have not improved for two tests. Reported performance is the one of the model taht minimizes $(\text{visualLoss} \times \lambda + \text{bitCost})$ on the validation set at the end of training. Base models are trained for six million steps at $\lambda=4096$ with the mean squared error (MSE) loss. Smaller $\lambda$ values and MS-SSIM models are trained for four million steps starting from the base model with their LR and optimizer reset. All models use $\text{C}_\text{H}=192$ (hidden layers channels) and $\text{C}_\text{L}=256$ (output channels) such that a single base model is needed for each prior configuration.
The training and validation dataset is made of free-license images from Wikimedia Commons \cite{commons}; mainly ``Category:Featured pictures on Wikimedia Common'' which consists of 13928 images of the highest quality. The images are cropped into $1024^2$ pixels patches on disk to speed up further resizing, then they are resized on-the-fly by a random factor down to $256^2$ pixels during training. A batch size of 4 patches is used. The kodak set \cite{kodak} is used as a validation set and the CLIC professional test dataset \cite{clictest} is used for testing.

\begin{figure}
\centering
\includegraphics[width=0.9\linewidth]{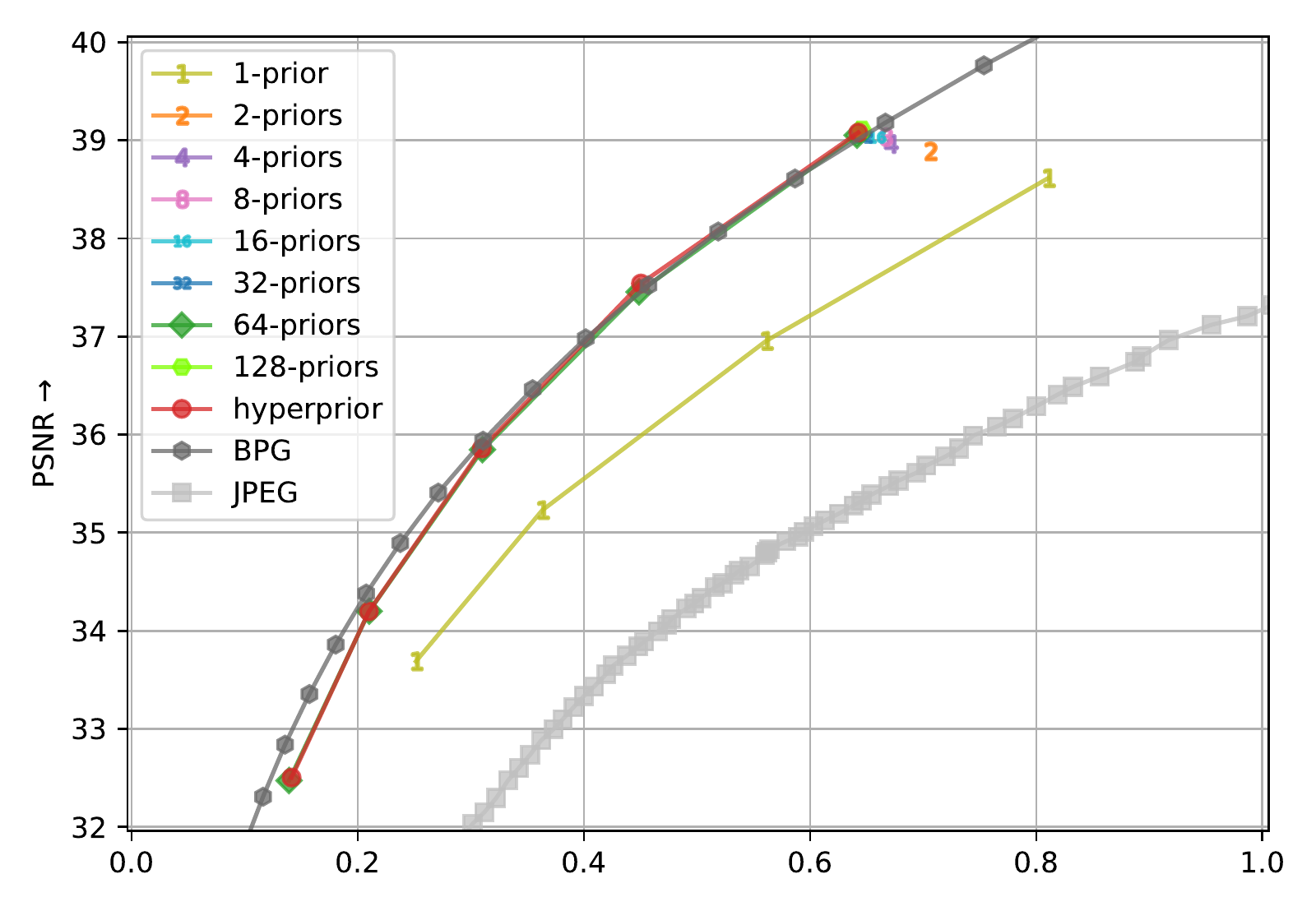}

\includegraphics[width=0.9\linewidth]{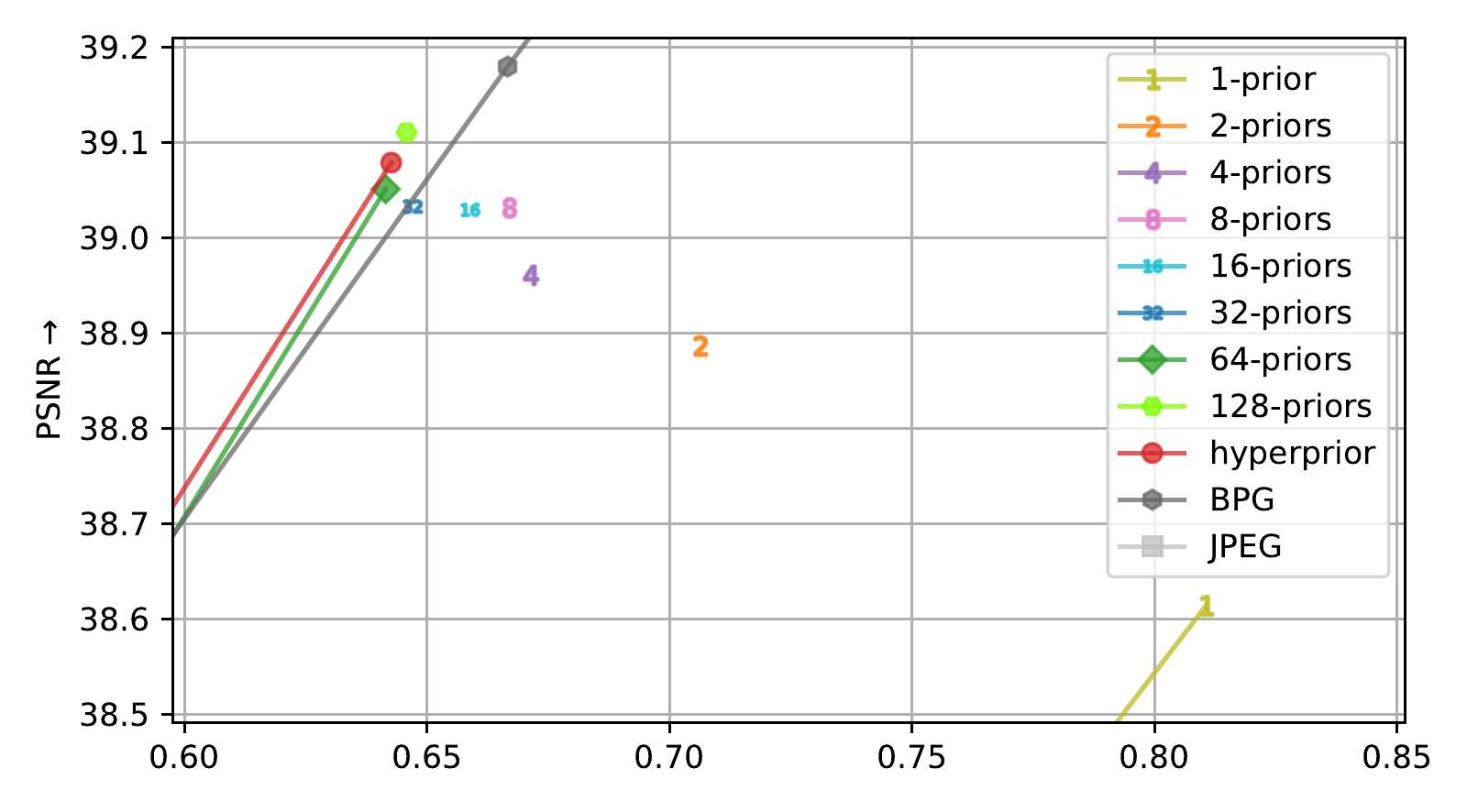}

\includegraphics[width=0.9\linewidth]{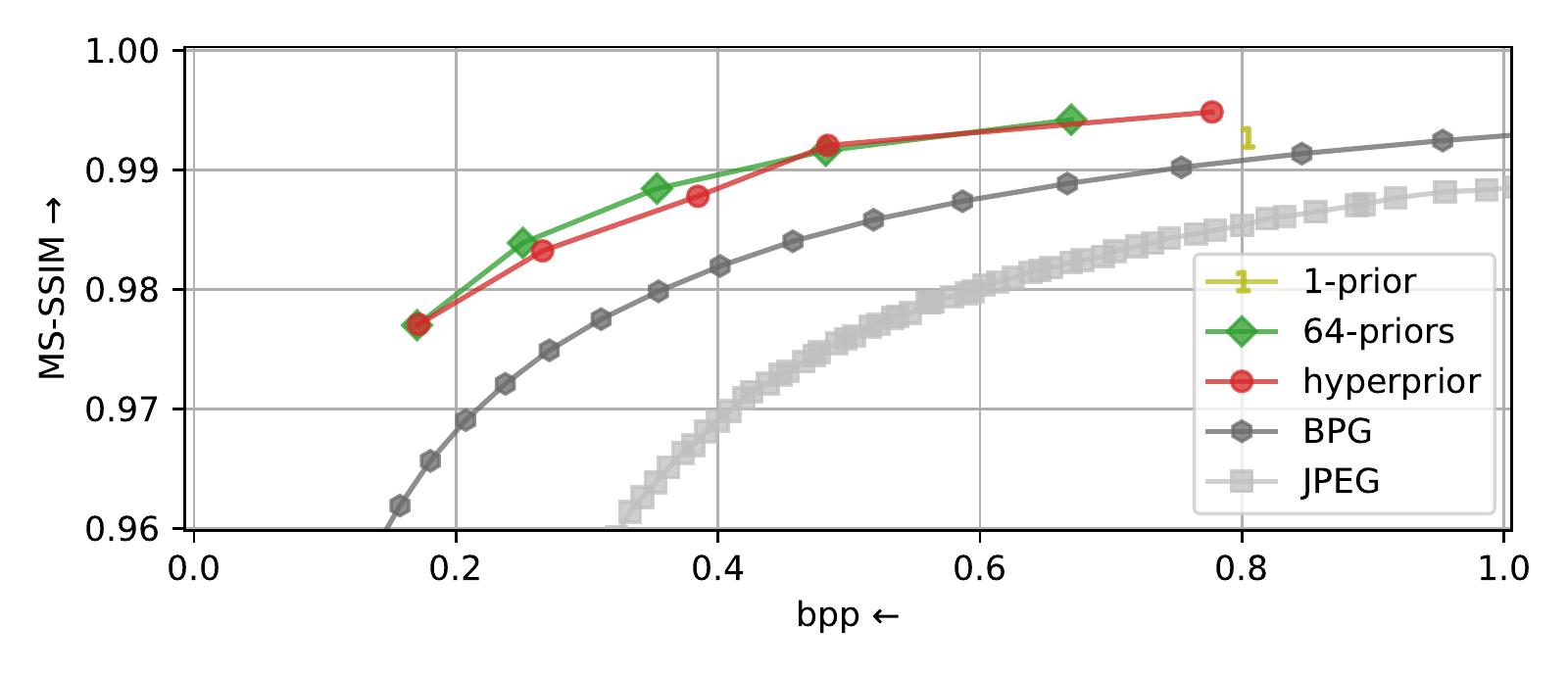}

\caption{Top: PSNR RD curve of a 64-priors model on the CLIC pro. test set, compared with the HP model \cite{balle2018}, and the BPG and JPEG codecs. Middle : Zoom in on models with 1, 2, 4, 8, 16, 32, and 64 priors. 
Bottom: MS-SSIM RD curve.}
\label{fig:bpp-psnr-clicpro}

\end{figure}

\begin{figure}
\centering
\includegraphics[width=\linewidth]{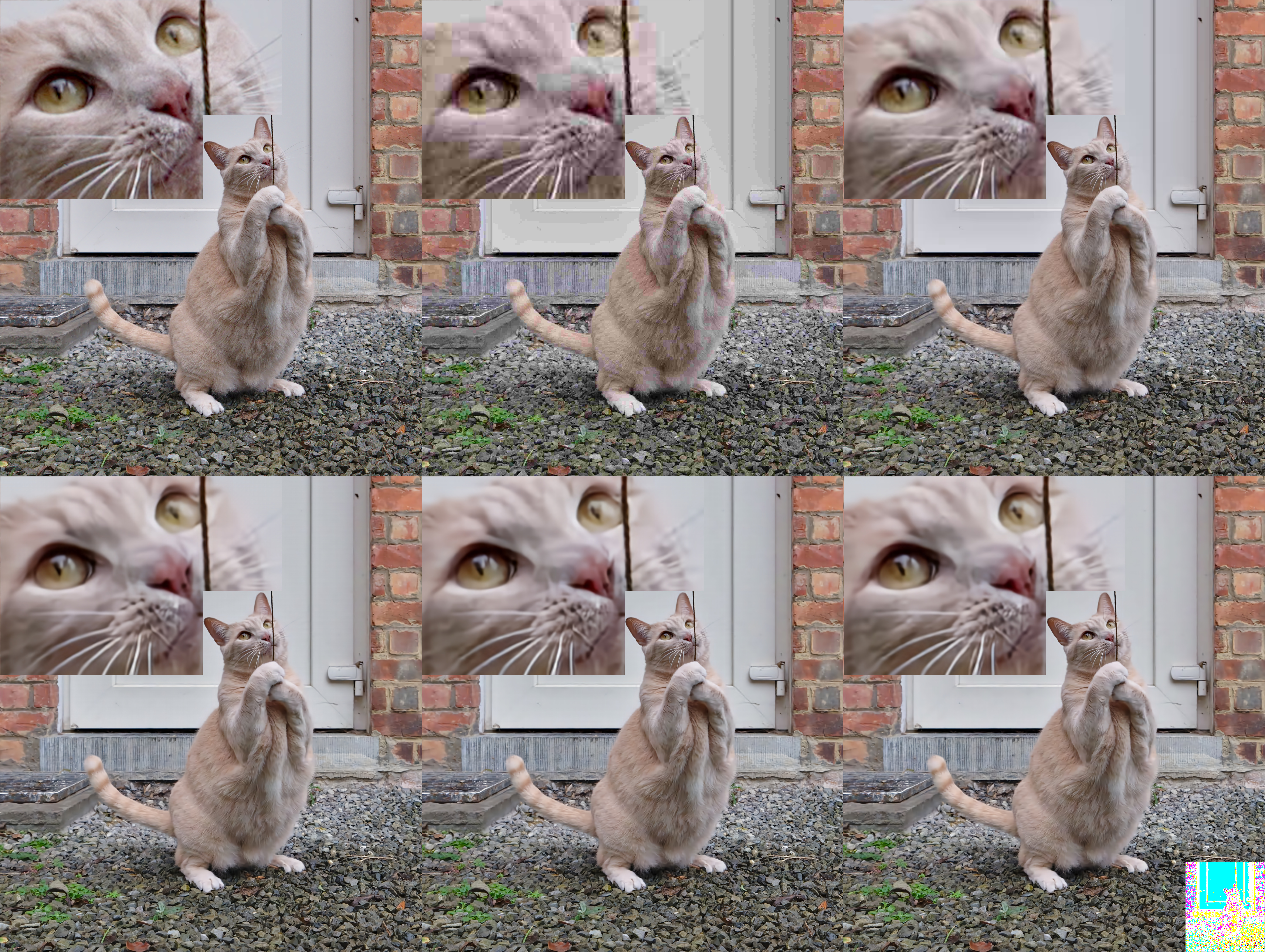}
\caption{Visual comparison of Larry the cat \cite{commonstestphotographs} compressed with learned ($\lambda=512$) and conventional methods. Top-left: uncompressed, top-middle: JPEG (PSNR: 29.3, 0.224 bpp), top-right: BPG (PSNR: 32.9, 0.217 bpp), bottom-left: 1-prior (PSNR: 32.4, 0.252 bpp), bottom-middle: hyperprior (PSNR: 32.8, 0.217 bpp), bottom-right: 64-priors/ours (PSNR: 32.9, 0.218 bpp)}
\label{fig:visualcomp}
\end{figure}

The RD curve of our ``multiprior'' model is compared with that of the HP model \cite{balle2018}, which is trained from scratch using Liu Jiaheng's PyTorch implementation \cite{ptcompression,liujiaheng}. 
Liu Jiaheng's code differs slightly from the paper's definition \cite{balle2018} in that a Laplace distribution is used in place of the normal distribution to stabilize training. Complexity is measured as the number of GMac (billion multiply-accumulate operation) using the ptflops counter \cite{ptflops} and the number of memory lookup operations is calculated manually.

\subsection{Results}

The PSNR RD curve measured on the CLIC professional test set \cite{clictest} is shown on top of \autoref{fig:bpp-psnr-clicpro}. The performance of a 64-priors model is in line with that of the HP model 
: they both perform slightly better than BPG at high bpp, and achieve significantly better RD than the single-prior model. 
In the middle, the RD value at $\lambda=4096$, the highest bitrate, is shown for 1, 2, 4, 8, 16, 32, 64, and 128 prior distributions. 128-priors offer marginal gains and costs an increased training time (1.5) and encoding time. 
MS-SSIM performance of fine-tuned models is shown in the bottom of \autoref{fig:bpp-psnr-clicpro}; the 64-priors model still performs similarly to \cite{balle2018}, and both learned compression models benefit from this more perceptual metric compared with traditional codecs. A visual comparison of images compressed with the MSE loss ($\lambda=512$) and the equivalent bitrate settings in conventional codecs is shown in \autoref{fig:visualcomp}.


\begin{table}[]
\small
\centering
\caption{Complexity of the HP model \cite{balle2018}) compared to Manypriors (ours), expressed in GMac for the neural network parts and number of memory lookup operations (* or parametrized Laplace CDF generations in full-precision) for the CDF tables generation, to process a 4K image.}
\label{tab:complexity}
\resizebox{1.\linewidth}{!}{
\begin{tabular}{lll|r|r|r}
($\downarrow$)                                                       &                                &                   & \multicolumn{1}{l|}{{\ul Hyperprior}} & \multicolumn{1}{l|}{{\ul Manypriors}} & \multicolumn{1}{l}{{\ul ratio MP$\div$HP}} \\ \hline
\multicolumn{1}{l|}{\multirow{7}{*}{{\ul \textbf{Encoding}}}} & \multirow{4}{*}{{\ul GMac}}    & main encoder      & {\ul 769.82}                          & {\ul 769.82}                          &                                                       \\
\multicolumn{1}{l|}{}                                              &                                & hyper encoder     & 23.75                                 &                                       &                                                       \\
\multicolumn{1}{l|}{}                                              &                                & hyper decoder     & 23.86                                 &                                       &                                                       \\ \cline{3-6} 
\multicolumn{1}{l|}{}                                              &                                & {\ul total}       & {\ul 817.43}                          & {\ul 769.82}                          & 0.942                                                 \\ \cline{2-6} 
\multicolumn{1}{l|}{}                                              & \multirow{3}{*}{{\ul Lookups}} & indices           & {\ul }                                & 530.84 M                              &                                                       \\
\multicolumn{1}{l|}{}                                              &                                & CDF               & 829.44 K *                            & 32.400 K                              &                                                       \\ \cline{3-6} 
\multicolumn{1}{l|}{}                                              &                                & {\ul total}       & {\ul 829.44 K *}                      & {\ul 530.87 M}                        & $\text{N}_\text{CDF} =64$                                          \\ \hline
\multicolumn{1}{l|}{\multirow{4}{*}{{\ul \textbf{Decoding}}}} & \multirow{3}{*}{{\ul GMac}}    & hyper decoder     & {\ul 23.154}                          & {\ul }                                &                                                       \\
\multicolumn{1}{l|}{}                                              &                                & main decoder      & 769.60                                & 769.60                                &                                                       \\ \cline{3-6} 
\multicolumn{1}{l|}{}                                              &                                & {\ul total}       & {\ul 792.75}                          & {\ul 769.60}                          & 0.971                                                 \\ \cline{2-6} 
\multicolumn{1}{l|}{}                                              & {\ul Lookups}                  & {\ul CDF (total)} & {\ul 829.44 K *}                      & {\ul 32.400 K}                        & ${1 \over \text{C}_\text{L}} = 0.004$      
\end{tabular}
}
\end{table}


Computational complexity of our Manypriors has been compared to the one of the HP model \cite{balle2018}). This complexity is expressed in GMac for the neural network parts and number of memory lookup operations. It is summarized in \autoref{tab:complexity}.
The lack of a HP AE saves 3 \% to 6 \% GMac, depending on whether only the HP decoder (image decoding) or the whole HP codec (image encoding) is used. 
Decoding with the Manypriors scheme is greatly simplified compared to \cite{balle2018} because the CDF tables generation process takes the optimal indices stored as side-information and looks up one static CDF table per latent spatial dimension, that is $\text{C}_\text{L}$ (typically 256) fewer lookups than with a HP.
 During encoding, the Manypriors scheme must lookup every latent variable with every CDF table in order to determine the most cost effective CDF tables. This results in $\text{N}_\text{CDF}$ (typically 64) times more lookup operations than the HP scheme overall, although these lookup operations are relatively cheap because only two values are needed (variable$\pm$0.5), whereas each CDF table lookup in \cite{balle2018} returns $\text{L}$ probabilities. Moreover, it is challenging to make an accurate CDF LUT for the HP scheme, because quantizing the distribution scale parameter reduces the accuracy of the resulting CDFs, negatively impacting the bitrate. This challenge is exacerbated when the distribution has multiple parameters \cite{minnen} or a mixture of distributions \cite{gmm} is used. 
In Figure \ref{fig:bpp-psnr-clicpro},  
LUT are replaced by accurate but complex Laplace distribution computation for the HP scheme in order to maximize the reported RD performance. 

Time complexity is measured for every step on CPU, where it can be reliably profiled due to synchroneous execution. It is summarized in \autoref{tab:timingtbl} with the following distinct sub-categories: NN (neural network) is the time spent in the AE, CDF generation is the time spent building the CDF tables for a specific image, and entropy is the bitstream generation. All operations are done using the PyTorch framework in python, except for entropy encoding which makes use of the torchac range coding library \cite{torchac,torchac-code}, written in C++, and the prior indices are compressed using the LZMA library \cite{lzma}. The total encoding time of the 64-priors model is 0.32 time that of the HP model and the decoding time is 0.42 times that of the HP model. The timing is more significant when it is broken down by sub-category because each component has a different response time depending on the hardware (and software) architecture in place. The AE (``NN'') encoding time is 0.90 that of the HP scheme and decoding time is 0.95 time as much as the HP. Both the hyper-encoder and hyper-decoder are called during encoding, thus it appears that each part of the HP sub-network costs 5 \% of the AE time. The time taken to build the CDF tables for the HP model was measured both by estimating the per-variable Laplace distributions (``full-precision'') and with a quantized scale parameter LUT. In any case, finding the best indices of a 64-priors model appears to be relatively inexpensive and the total CDF tables generation time is only 0.17 to 0.48 that of the HP model (depending on whether the HP model uses full-precision or LUT) for encoding. During decoding, the 64-priors model spends 0.05 to 0.14 as much time building the CDF tables as the HP model, because the optimal CDF table indices have already been determined during encoding and they are included in the bitstream.

\begin{table}[]
\small
\centering
\caption{Breaking down the image encoding and decoding time, in seconds. 
Image: 4.5 MP snail \cite{commonstestphotographs}. CPU: AMD Ryzen 7 2700X. Time avg. of 50 runs.}
\label{tab:timingtbl}
\resizebox{1\linewidth}{!}{
\begin{tabular}{ll|rr|r}
($\downarrow$)                                 & {\ul \textbf{}}               & \multicolumn{1}{l}{\begin{tabular}[c]{@{}l@{}}Hyperprior\\ (Ballé2018)\end{tabular}} & \multicolumn{1}{l|}{\begin{tabular}[c]{@{}l@{}}64-priors\\ (ours)\end{tabular}} & \multicolumn{1}{l}{\begin{tabular}[c]{@{}l@{}}ratio\\ (ours$\div$HP)\end{tabular}} \\ \hline
\multirow{5}{*}{{\ul \textbf{Encoding}}} & NN encode: main + hyperprior                & 3.81 + 0.41                                                                                 & 3.79       + 0.00                                                                      & 0.90                                                                                           \\ \cline{2-4}
\cline{2-4}
                                               & entropy encode, main + hyperprior           & 0.15 + 0.02                                                                                & 0.15   + 0.00                                                                          &                                                                                           \\ \cline{2-5} 
                                               & { \textbf{CDF}: select indices + gather tables}     & 0.00 + \begin{tabular}[c]{@{}r@{}}FP: 15.95\\ LUT: 5.66\end{tabular}                          & 1.90 + 0.81                                                                             & \textbf{\begin{tabular}[c]{@{}r@{}}FP: 0.17\\ LUT: 0.48\end{tabular}}                                \\ \cline{2-5} 
                                               & {\ \textbf{encode (total)}} & \begin{tabular}[c]{@{}r@{}}FP: 20.33\\ LUT: 10.04\end{tabular}                & 6.65                                                                    & \textbf{\begin{tabular}[c]{@{}r@{}}FP: 0.32\\ LUT: 0.66\end{tabular}}                       \\ \hline
\multirow{5}{*}{{\ul \textbf{Decoding}}} 
                                               & { NN decode : main + hyperprior}         & 10.66 + 0.34                                                                                & 10.50                                                                            & 0.95                                                                                      \\ \cline{2-5} 
                                               & { \textbf{CDF} : gather tables}       & \begin{tabular}[c]{@{}r@{}}FP: 15.95\\ LUT: 5.66\end{tabular}                          & 0.81                                                                             & \textbf{\begin{tabular}[c]{@{}r@{}}FP: 0.05\\ LUT: 0.14\end{tabular} }                               \\ \cline{2-5} 
                                               & { entropy decode : main + hyperprior}    & 0.24 + 0.02                                                                                 & 0.24                                                                             & 0.92                                                                                      \\ \cline{2-5} 
                                               & { \textbf{decode (total)}}   & \begin{tabular}[c]{@{}r@{}}FP: 27.21\\ LUT: 16.92\end{tabular}                & 11.54                                                                   & \textbf{\begin{tabular}[c]{@{}r@{}}FP: 0.42\\ LUT: 0.68\end{tabular}}                      
\end{tabular}
}

\end{table}

\section{Conclusion}

Convolutional autoencoders trained for compression are optimized for both rate and distortion. Rate is estimated with a cumulative probability model, which in turns generates a CDF for every latent variable to be encoded. A single CDF per latent channel 
is not sufficient to capture the statistics at the output of the encoder, nor to allow the encoder to express a wide variety of features. To support multiple statistics, the hyperprior \cite{balle2018} parametrizes a standard distribution, 
but this introduces a great deal of complexity in the entropy coding stage because the CDF differs for every latent variable to be encoded. The proposed method uses multiple prior distributions working as a competition of experts to capture the relevant features which they specialize on. This approach is advantageous because the learned CDF tables are stored in a static LUT once training is finished, and a model trained with 64 prior distributions performs with a similar RD as one trained with a HP sub-network. 
Moreover, a learned CDF table includes the CDF for all channels in the latent code. Hence, accessing the CDF table for a spatial location provides the CDF for each of its channels and the number of lookups is reduced to the number of latent spatial locations. In our experiments, CDF tables generation in the encoding step takes 0.17 to 0.48 as much time with a 64-priors model as it does with the HP model (depending on the precision of the HP model). This ratio is lowered to 0.05 to 0.14 during decoding because the prior indices have already been determined during the encoding.

\section{Acknowledgements}

This research has been funded by the Walloon Region. Computational resources have been provided by the supercomputing facilities of the Université catholique de Louvain (CISM/UCL) and the Consortium des Équipements de Calcul Intensif en Fédération Wallonie Bruxelles (CÉCI) funded by the Fond de la Recherche Scientifique de Belgique (F.R.S.-FNRS) under convention 2.5020.11 and by the Walloon Region.

{\small
\bibliographystyle{ieee_fullname}
\bibliography{multipriors}
}

\end{document}